\documentclass[a4paper, 12pt]{article}

\usepackage{graphicx}
\usepackage{url}
\usepackage{amsbsy,amssymb} 
\usepackage{amsmath}
\usepackage{abstract}
\usepackage{hyperref}
\usepackage{color}
\begin{document}

%%%%%%%%%%%%%%%%%%%%%%%%%%%%%%%%%%%%%%%%%%%%%%%%%%%%%%%%%%%%%%%%%%%%%%%%%%%%%%%
{\LARGE\centering{\bf{The design and the performance of stratospheric mission in the search for the Schumann resonances}}}
%%%%%%%%%%%%%%%%%%%%%%%%%%%%%%%%%%%%%%%%%%%%%%%%%%%%%%%%%%%%%%%%%%%%%%%%%%%%%%%%%
%authors
%%%%%%%%%%%%%%%%%%%%%%%%%%%%%%%%%%%%%%%%%%%%%%%%%%%%%%%%%%%%%%%%%%%%%%%%%%%%%%%%%

\begin{center}
\sf{ Arkadiusz Papaj$^{a}$\footnote{arkadiusz.papaj@gmail.com}, Piotr Weszka$^{a}$\footnote{piotrweszka@gmail.com }, Marcin Boche\'{n}ski$^{a}$\footnote{bochenski.marcin@outlook.com}, Mateusz Micha\l{}ek$^{a}$\footnote{mateusz.michalek@pk.edu.pl}, Andrzej Ku\l{}ak$^{b}$\footnote{kulak@oa.uj.edu.pl}, Agata Ko\l{}odziejczyk$^{c}$\footnote{maria.kolodziejczyk@esa.int}, Matt Harasymczuk$^{d,c}$\footnote{matt@harasymczuk.pl}, Pawe\l{} Karbowniczek$^{a}$\footnote{pkarbowniczek@pk.edu.pl}, Aleksandra \L{}awrynowicz$^{e}$\footnote{olalawrynowicz@gmail.com}, Joanna Ku\'{z}ma $^{f}$\footnote{joanna.kuzmaa@gmail.com}, Tomasz Brol\footnote{t.brol@ssh.gliwice.pl}, Rados\l{}aw A. Kycia$^{a}$\footnote{Corresponding author: kycia.radoslaw@gmail.com}}
\end{center}

\medskip
\small{
\begin{center}
$^{a~}$Tadeusz Ko\'{s}ciuszko Cracow University of Technology, \\ Warszawska 24, 31-155 Krak\'{o}w, Poland. \\
$^{b~}$AGH University of Science and Technology, \\Faculty of Computer Science, Electronics and Telecommunications, \\ Institute of Electronics, \\ Al. A. Mickiewicza 30, Pawilon D-17,\\ 30-059 Krak\'{o}w, Poland.\\
$^{c~}$ European Space Agency \\ Advanced Concepts Team \\ ESTEC Keplerlaan 1, PO Box 299, \\NL-2200 AG Noordwijk, The Netherlands.\\ 
$^{d~}$ Warsaw University \\ Krakowskie Przedmie\'{s}cie 26/28, 00-927 Warszawa, Poland. \\
$^{e~}$ Pomorski Park Naukowo-Technologiczny Gdynia  \\ Al. Zwyciestwa 96-98, \\ 81-451 Gdynia, Poland.\\
$^{f~}$ Wroc³aw University of Science and Technology, \\ Faculty of Chemistry, \\  Norwida 4/6, 50-373 Wroc\l{}aw, Poland.\\
\end{center}
}
\bigskip

PACS 2010: 94.05.Rx, 94.05.Sd, 94.20.ws

%%%%%%%%%%%%%%%%%%%%%%%%%%%%%%%%%%%%%%%%%%%%%%%%%%%%%%%%%%%%%%%%%%%%%%%%%%%%%%
%absract
%%%%%%%%%%%%%%%%%%%%%%%%%%%%%%%%%%%%%%%%%%%%%%%%%%%%%%%%%%%%%%%%%%%%%%%%%%%%%%
\begin{abstract}
\noindent
The technical details of a balloon stratospheric mission that is aimed at measuring the Schumann resonances are described. The gondola is designed specifically for the measuring of faint effects of ELF (Extremely Low Frequency electromagnetic waves) phenomena.  The prototype met the design requirements. The ELF measuring system worked properly for entire mission; however, the level of signal amplification that was chosen taking into account ground-level measurements was too high. Movement of the gondola in the Earth magnetic field induced the signal in the antenna that saturated the measuring system. This effect will be taken into account in the planning of future missions. A large telemetry dataset was gathered during the experiment and is currently under processing. The payload consists also of biological material as well as electronic equipment that was tested under extreme conditions. 
\end{abstract}

%%%%%%%%%%%%%%%%%%%%%%%%%%%%%%%%%%%%%%%%%%%%%%%%%%%%%%%%%%%%%%%%%%%%%%%%%%%%
%Introduction
%%%%%%%%%%%%%%%%%%%%%%%%%%%%%%%%%%%%%%%%%%%%%%%%%%%%%%%%%%%%%%%%%%%%%%%%%%%%
\section{Introduction}
%%%%
ELF are the waves that are usually connected with natural atmospheric phenomena. They are defined to have frequency from $3Hz$ to $3kHz$. Its observation can be affected by industrial activity that disturb the measurements - the most common is $50Hz$ radiation of electric network and radio transmissions. Therefore, the measuring device has to be designed in the way that it can discard of these strong disturbances.

One of the most important ELF phenomena are the Schumann resonances, which were predicted and measured in 50s of XX century (see \cite{SchumannMeasurment}) and the references therein. They occur as the Earth-ionosphere waveguide is constantly powered by the electromagnetic waves from lightning. They interfere giving characteristic spectrum of amplifications at frequencies $7.83Hz$ and higher harmonics $14.3Hz$, $20.8Hz$ etc. \cite{ElectricNatureOfStorm}. The frequency location of the resonances is connected with the parameters of the atmosphere therefore they can be used for measuring of their properties on Earth and other planets, e.g., in future missions on Mars \cite{MARS}, \cite{MARS2}. There are also some indications that the low frequency electromagnetic fields influence biological systems, however, there is no general description of this phenomena - see review in \cite{BiologicalSystemsELF}.

Stratospheric balloon missions are the most versatile, in the sense of cost - results optimisation, way of performing measurements in the environment closely connected to the cosmic space or in the higher layers of atmosphere. As the ascending and descending phases are slower than in rocket carriers the gathered datasets can be large. It is therefore natural to use this platform to measure the resonances, as it was suggested in \cite{MARS} in case of Mars. The first initial results of measuring the resonance were reported in \cite{BaloonSchumann}, therefore, it suggests that the proper design of balloon ELF mission is of great importance in atmospheric research and future cosmic exploration.

The paper is organized as follows. In the next section the overview of the electronic antenna system and then the design of the gondola will be presented. Next, the brief description of the mission will be outlined and conclusions from the first iteration will be presented.

%%%%%%%%%%%%%%%%%%%%%%%%%%%%%%%%%%%%%%%%%%%%%%%%%%%%%%%%%%%%%%%%%%%%%%%%%%%%%%%%%%%%%%%%%%%%%%%%%%%%%%%%%%%%%%%%%%%%%%%%%
\section{Antenna system}
%%%%%%%%%%%%%%%%%%%%%%%%%%%%%%%%%%%%%%%%%%%%%%%%%%%%%%%%%%%%%%%%%%%%%%%%%%%%%%%%%%%%%%%%%%%%%%%%%%%%%%%%%%%%%%%%%%%%%%%%%
The Schumann resonance has two components - the electric one which is vertical and the magnetic one. The experiment was aimed at measurement of the first case.

The construction of antenna as the standard dipole \cite{AntennaTheory} is unsuitable for balloon experiments due to the long wavelengths. The most appropriate choice, as the space and mass of package is constrained by the avionic law, is the short dipole active antenna of length of a few centimetres comparing to the wavelengt of hundreds of kilometres of ELF waves. In the field of ELF wave it behaves like the electromotive source with negligible resistance and inductance. Therefore, it has to be connected with amplifier with large input impedance and small capacity. The most optimal length for balloon missions is $20 cm$, which, basing on the ground level measurements of the Schumann resonances, would generate output on the level of $90\mu V$, which results from the standard theory of short dipole of given length and estimated value of electric field on the ground level.

The scheme of the amplifier system is a small modification of the design from \cite{ELFAntenna} called ELA 1 with passive antenna. For summary see also \cite{ELFAntenna2}. This design was used to observe ELFs on the ground \cite{ELFAKulak1}, \cite{ELFAKulak2}, \cite{ELFAKulak3}, \cite{ELFAKulak4}. It was supplied with Chebyshev filter reducing aliasing. The output was connected to ADC described in the next section.

In the ground tests the dominating signal $50Hz$ of electric power was visible, that showed the antenna system worked. In this design the induction of signal in the antenna by movement in Earth magnetic field was not taken into account as the effect depends the parameters of flight and wind.

%%%%%%%%%%%%%%%%%%%%%%%%%%%%%%%%%%%%%%%%%%%%%%%%%%%%%%%%%%%%%%%%%%%%%%%%%%%%%%%%%%%%%%%%%%%%%%%%%%%%%%%%%%%%%%%%%%%%%%%%%
\section{Data acquisition system}
%%%%%%%%%%%%%%%%%%%%%%%%%%%%%%%%%%%%%%%%%%%%%%%%%%%%%%%%%%%%%%%%%%%%%%%%%%%%%%%%%%%%%%%%%%%%%%%%%%%%%%%%%%%%%%%%%%%%%%%%%
The system of data acquisition consists of two computers for backup purposes:
\begin{enumerate}
 \item {RaspberryPi 3 B, 4-channel 12bit ADC converter - ADS1015, GPS and IMU(Inertial Measurement Unit) GY-801;}
 \item {Arduino Due, 16 bit ADC - ADS1115 and GPS;}
\end{enumerate}
The systems were charged by the TP-LINK TL-PB10400 power bank with 10400 mAh capacity. The power bank has two ports and one of them charged RaspberryPi and Arduino computers and the second one the antenna system. There was also the second battery (Colorovo PowerBox 6800mAh) connected to the YI Action 2 camera, which was also the device that was tested against low temperatures and extreme stratospheric conditions. 10400 mAh power supply was too large for 2h mission(as it was tested before the mission), however it was used in order to prevent effect of low temperature on capacity of chemical power sources. The data were saved on the fast input-output transaction SD cards.

The data frame format used in the first system was as follows
\begin{verbatim}
GPS: DATA
GPS: DATA
[ACCELERATION x,y,z] [GYROSCOPE x,y,z] [MAGNETIC FIELD x,y,z]
[
'TIME', ADC,
'TIME', ADC,
...
]
...
END
\end{verbatim}
where the first two lines are GPS data, then the data from IMU. Next part is the data from timer and corresponding ADC readout of $3000$ samples and finally END marker. The average sampling ratio was $300Hz$, which is sufficient to detect the Schumann resonances. The clock was synchronized with GPS at startup of the system before the start.

The data frame of the second system has the following format
\begin{verbatim}
 GPS
 ADC
 ...
 GPS
 ADC
 ...
 \end{verbatim}
where GPS stands for the data from GPS and ADC denotes the data from ADC. Arduino was not connected to RTC(Real Time Clock), therefore it was saving ADC data until the GPS did not interrupt, which ended the frame - the number of ADC readouts depends on the frequency of GPS interruptions. Its average sampling rate was $600-700Hz$ - the spread results from interrupt-driven design.

%%%%%%%%%%%%%%%%%%%%%%%%%%%%%%%%%%%%%%%%%%%%%%%%%%%%%%%%%%%%%%%%%%%%%%%%%%%%%%%%%%%%%%%%%%%%%%%%%%%%%%%%%%%%%%%%%%%%%%%%%
\section{Gondola}
\begin{figure}
\centering
 \includegraphics[width = 0.4\textwidth]{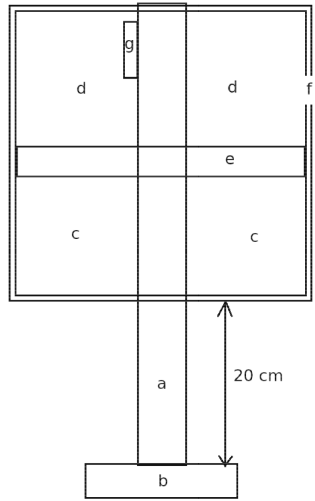}
 \caption{Schematic sketch of gondola(no real sizes): a - antenna(plastic pipe covered by metallic foil); b - protection of the antenna sharp end(XPS material); c- bottom compartment(for antenna system); d- top compartment(for acquisition system, telemetry, battery and camera); e - partition wall from XPS material; f- hole for camera; g - Inertial Measurement Unit;}
 \label{Fig:Gondola}
\end{figure}

%%%%%%%%%%%%%%%%%%%%%%%%%%%%%%%%%%%%%%%%%%%%%%%%%%%%%%%%%%%%%%%%%%%%%%%%%%%%%%%%%%%%%%%%%%%%%%%%%%%%%%%%%%%%%%%%%%%%%%%%%
Gondola was designed to met standards of aviation law. Its mass was $1.69kg$. The whole gondola was made from the pieces of XPS(Extruded Polyester) glued together. The outer layer was covered by aluminium foil in order to prevent electrostatic discharges which could disrupt ELF measurements. The sharp end of the antenna was protected by a piece of XPS. The Fig. \ref{Fig:Gondola} presents its cross section along the centre.
%%%%
%%%

\begin{figure}
\centering
 \includegraphics[width = 0.9\textwidth]{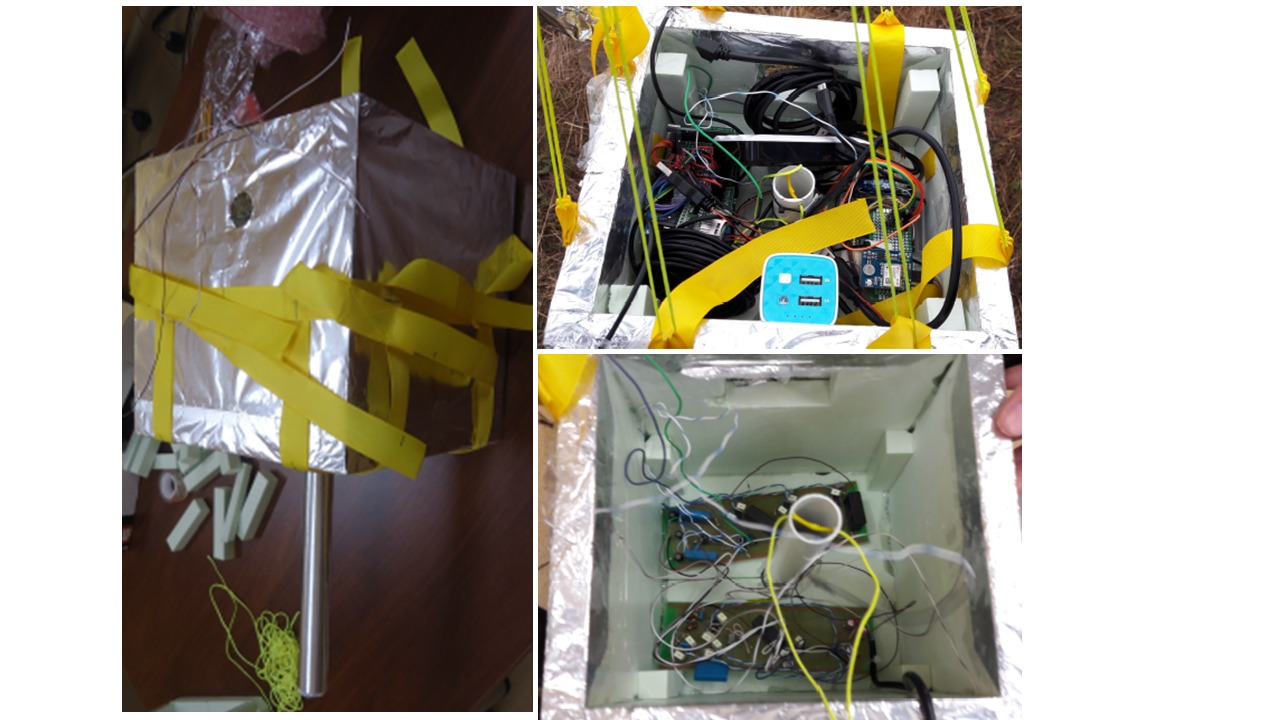}
 \caption{Capsule view - side view, top compartment, bottom compartment;}
 \label{Fig:GondolaPictures}
\end{figure}

The bottom isolated compartment was occupied by the antenna system. The upper part was occupied by the acquisition systems, GPS and battery. In addition APRS(Automatic Packet Reporting System) which allows to localize balloon on-line was also present. In the top part there was also a place for camera, which was placed for tests in stratospheric conditions - there was a hole for objective; see Fig. \ref{Fig:GondolaPictures}.
%%%%
%%%
Additional payload of biological samples was attached to the side walls of the gondola as well as to the long string and hang below the gondola.

The gondola was attached to the parachute by nylon strings and the parachute was connected with a balloon in such a way that in case it popped it automatically opens during fall.

For the experiment Hwoyee HY-1200 balloon model was selected as it is sufficient to reach $30km$ with payload of $2kg$ when filled with hydrogen gas.

In the next section the mission will be described.
%%%%%%%%%%%%%%%%%%%%%%%%%%%%%%%%%%%%%%%%%%%%%%%%%%%%%%%%%%%%%%%%%%%%%%%%%%%%%%%%%%%%%%%%%%%%%%%%%%%%%%%%%%%%%%%%%%%%%%%%%
\section{Mission}
\begin{figure}
\centering
 \includegraphics[width = 0.6\textwidth]{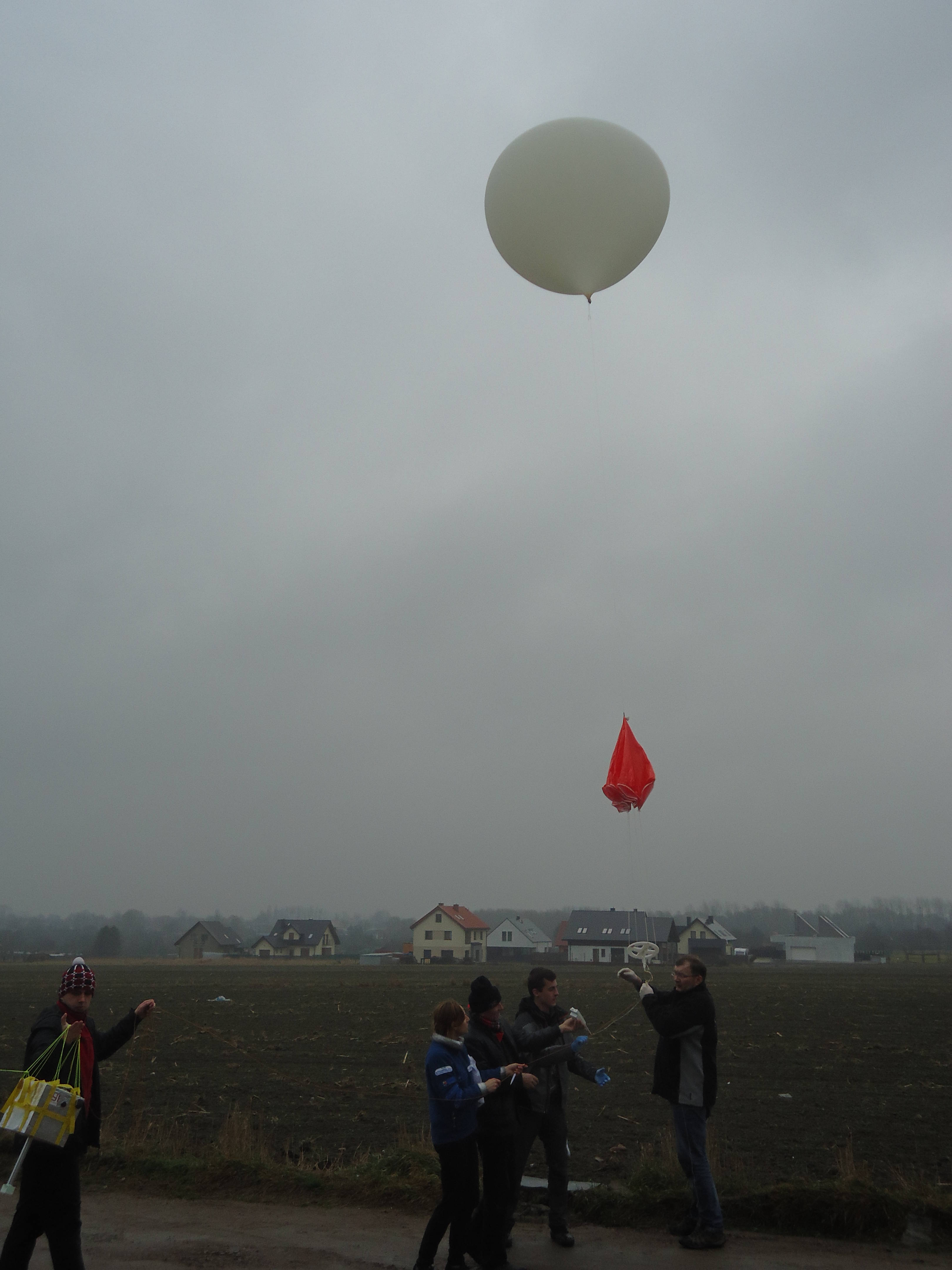}
 \caption{Start of the balloon.}
 \label{Fig:Start}
\end{figure}

\begin{figure}
\centering
 \includegraphics[width = 0.9\textwidth]{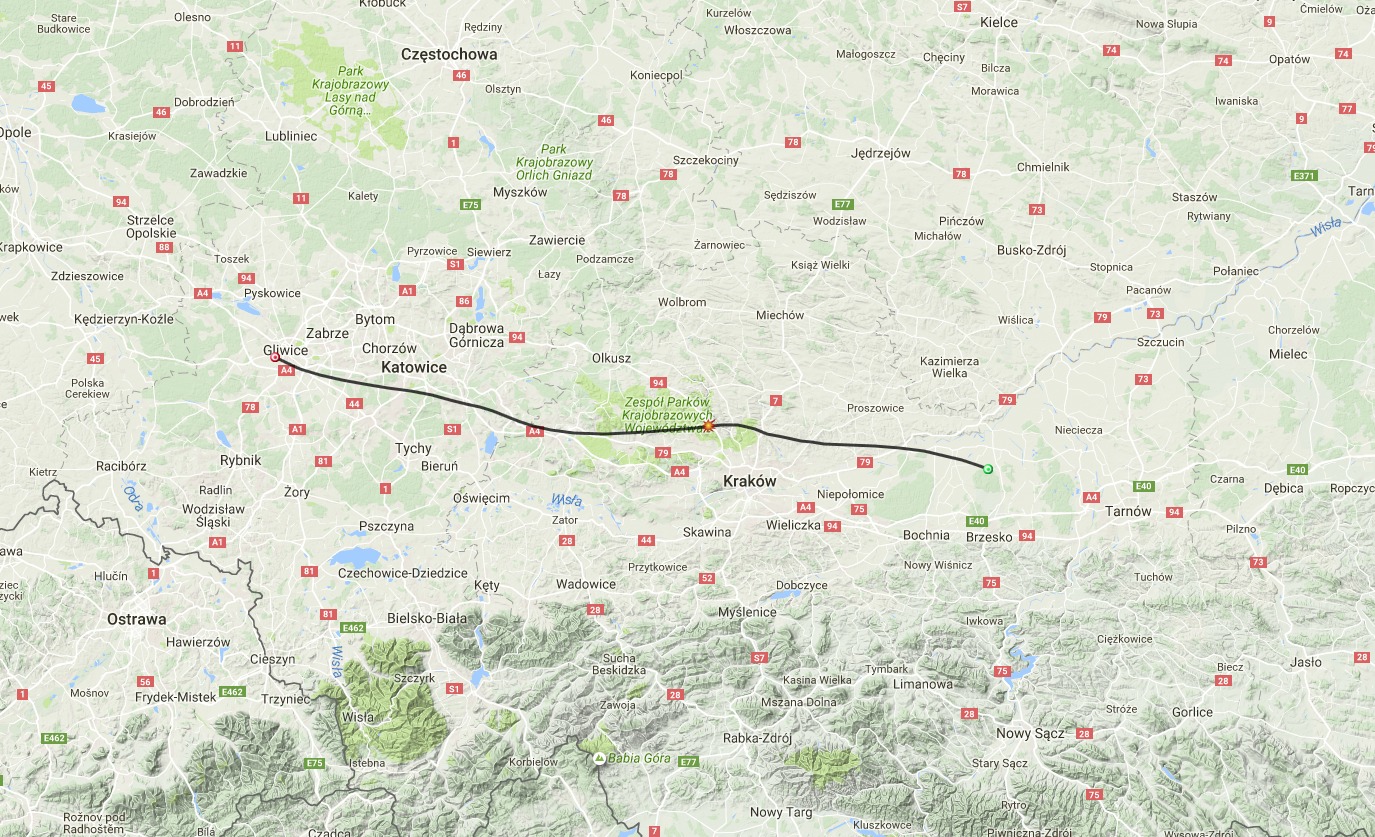}
 \caption{Trajectory of the balloon predicted according to the wind simulation on 24th of November from \cite{predykcja}.}
 \label{Fig:BaloonPathPrdiction24}
\end{figure}

\begin{figure}
\centering
 \includegraphics[width = 0.9\textwidth]{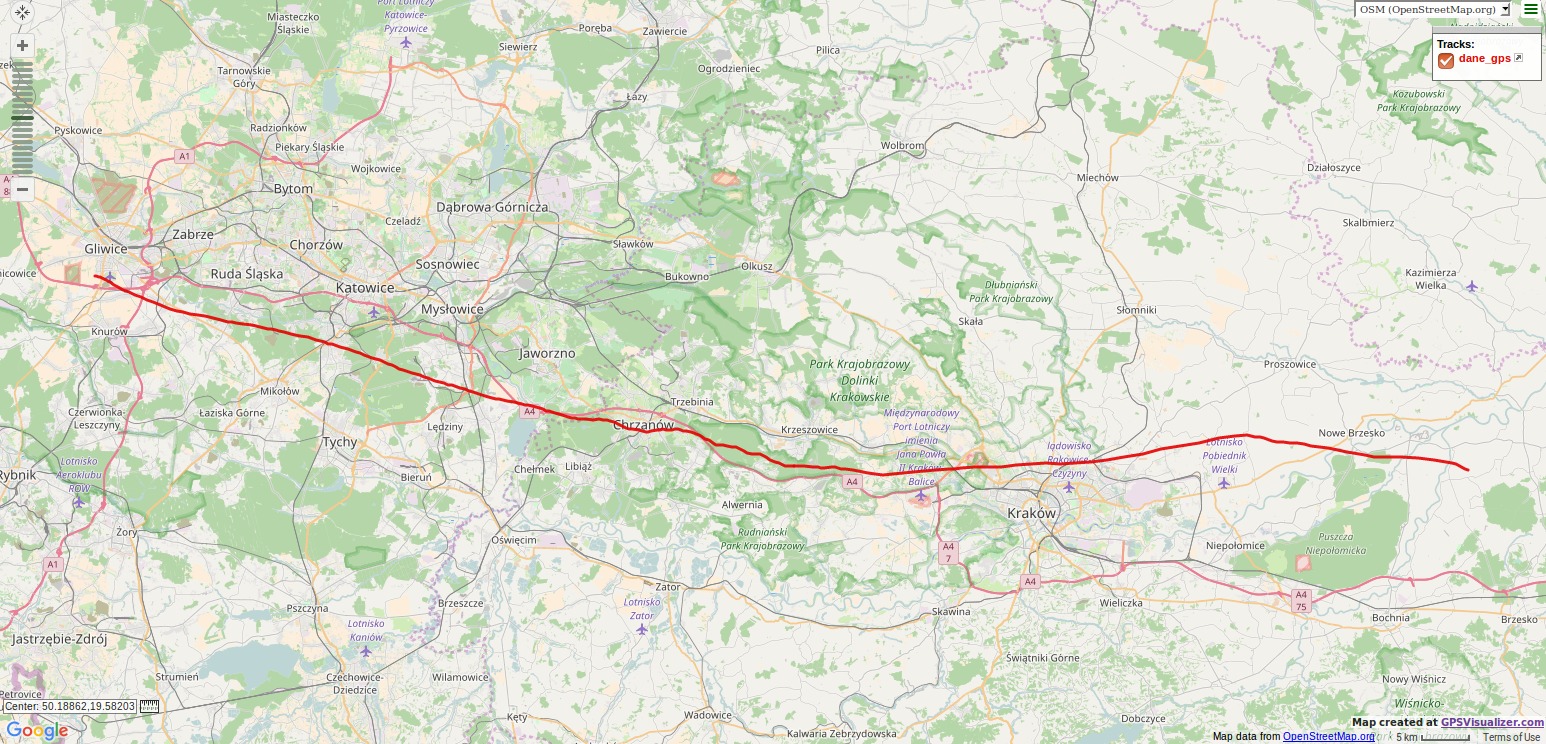}
 \caption{Trajectory of the balloon generated from GPS data from \cite{gpsVisualizer}.}
 \label{Fig:BaloonPath2}
\end{figure}

%%%%%%%%%%%%%%%%%%%%%%%%%%%%%%%%%%%%%%%%%%%%%%%%%%%%%%%%%%%%%%%%%%%%%%%%%%%%%%%%%%%%%%%%%%%%%%%%%%%%%%%%%%%%%%%%%%%%%%%%%
The mission started on 27 November 2016 at 9:18 am of CET, i.e. GMT(+1h) when the balloon was released near Gliwice, Poland, see Fig. \ref{Fig:Start}.
%%%%
%%%
The decision of start was preceded by the simulation of the trajectory using wind predictions at \cite{predykcja}. The prediction for the balloon trajectory shortly after the start of the balloon is presented in Fig. \ref{Fig:BaloonPathPrdiction24} and the path of flight from GPS data is presented in Fig. \ref{Fig:BaloonPath2}. 
%%%%
%%%
%%%%
%%%
The comparison of Fig. \ref{Fig:BaloonPath2} with the simulation made on 24 of November on Fig. \ref{Fig:BaloonPathPrdiction24} indicates that the simulation quite well agrees with the real path.

\begin{figure}
\centering
 \includegraphics[width = 0.9\textwidth]{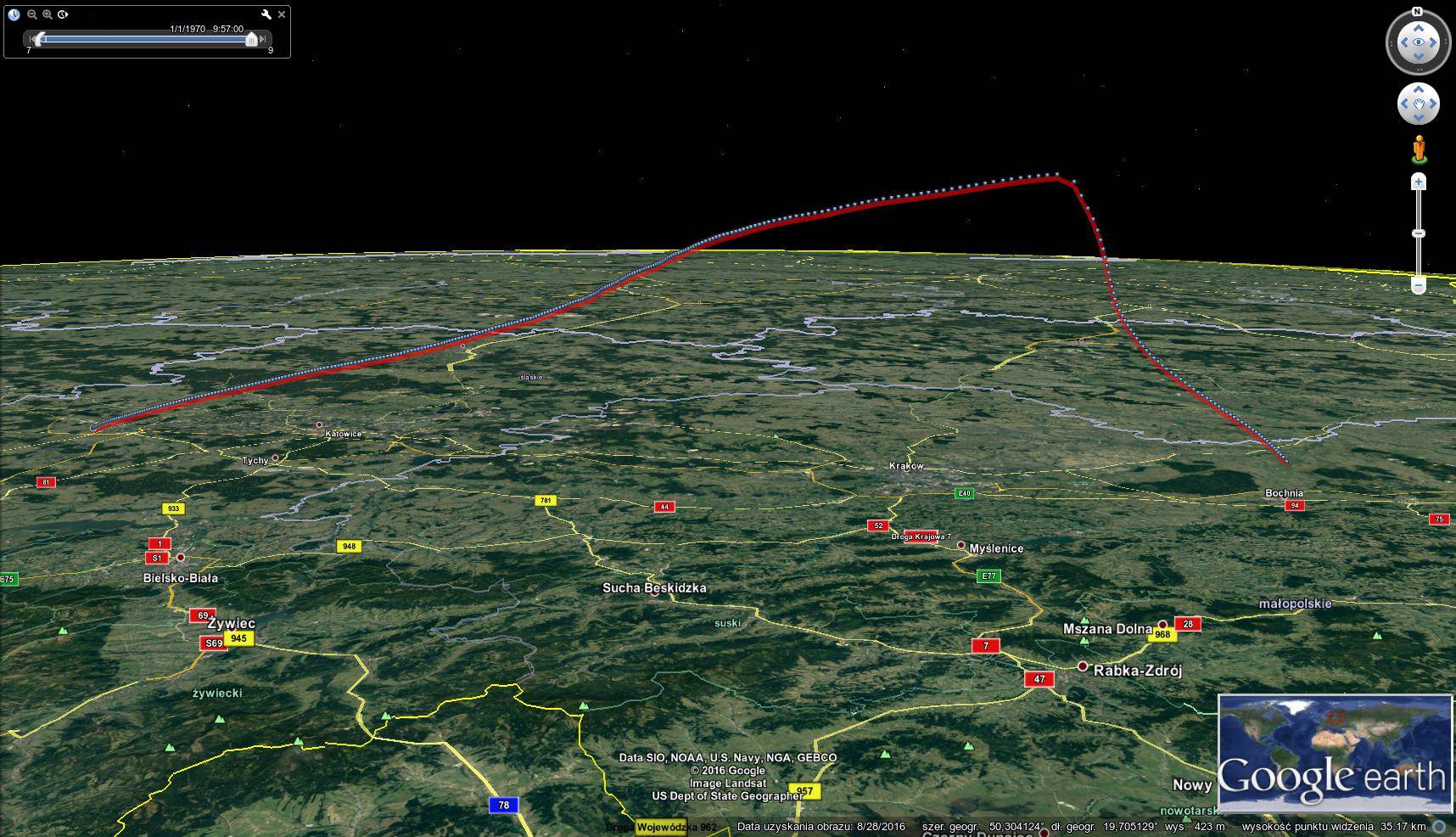}
 \caption{The path of the balloon from APRS data visualized in Google Earth service \cite{GOOGLEEarth}.}
 \label{Fig:BaloonPath}
\end{figure}

In addition height profile of the path of the balloon from APRS data are presented in Fig. \ref{Fig:BaloonPath}.
%%%%
%%%

\begin{figure}
\centering
 \includegraphics[width = 0.9\textwidth]{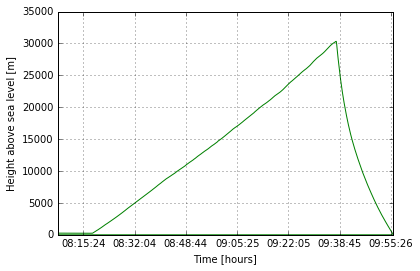}
 \caption{Height profile for GPS data. Time is given with the respect to GMT.}
 \label{Fig:BaloonHeight}
\end{figure}

\begin{figure}
\centering
 \includegraphics[width = 0.9\textwidth]{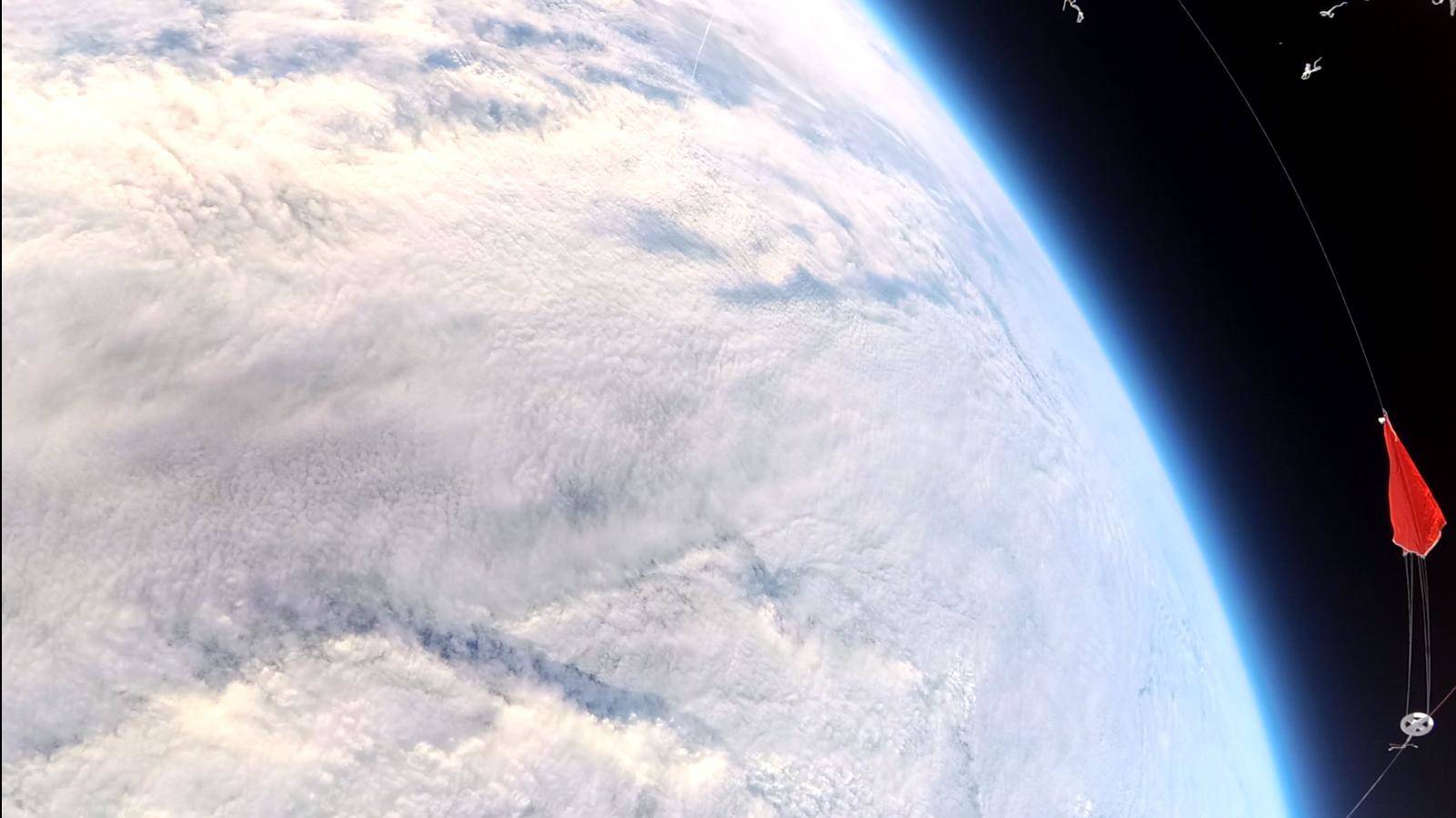}
 \caption{Balloon popping. The opening of the parachute (red color) and the remnants of the balloon are visible. The large cloud cover and the atmosphere layer that gradually passes to the space can be noticed.}
 \label{Fig:BaloonPopped}
\end{figure}

The height plot for the data from GPS is presented in Fig. \ref{Fig:BaloonHeight}. It can be seen that the balloon had constant vertical speed during ascending(line), however after it popped its vertical velocity was large(almost horizontal part of the trajectory) until it started to decelerate when the parachute was slowly opening. More detailed analysis of the flight can describe dynamics of the atmosphere and gondola-balloon system and it will be presented in a separate paper.
%%%%
%%%
The balloon popped at the attitude of $30km$ above see level and during the descend the parachute opened as it can be seen in Fig. \ref{Fig:BaloonPopped}.
%%%%
%%%

The flight lasted 2 hours. The gondola travels $135km$ from the starting to the landing point and the lowest temperature which it was exposed to was $-55^{\circ}C$. 

Additional information from the mission, including photos, are available on \cite{AdditionalMaterials}.

The preliminary results of experiment will be described in the next section.

%%%%%%%%%%%%%%%%%%%%%%%%%%%%%%%%%%%%%%%%%%%%%%%%%%%%%%%%%%%%%%%%%%%%%%%%%%%%%%%%%%%%%%%%%%%%%%%%%%%%%%%%%%%%%%%%%%%%%%%%%
\section{Preliminary results}
\begin{figure}
\centering
 \includegraphics[width = 0.9\textwidth]{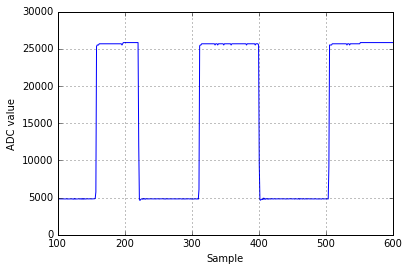}
 \caption{Saturation of the ELF system. The system oscillates between low and high state.}
 \label{Fig:Saturation}
\end{figure}

%%%%%%%%%%%%%%%%%%%%%%%%%%%%%%%%%%%%%%%%%%%%%%%%%%%%%%%%%%%%%%%%%%%%%%%%%%%%%%%%%%%%%%%%%%%%%%%%%%%%%%%%%%%%%%%%%%%%%%%%%
Preliminary analysis of the data from the Schumann resonances measurement system indicated that the system was saturated for the entire mission as it was presented in Fig. \ref{Fig:Saturation}. 
%%%%
%%%
This indicates that the level of amplification was too high. Therefore the redesigning of this part of the system is required before the next mission. The excitement was caused by the move in the earth magnetic field as it was tested after the mission on the ground tests. 

The analysis of data from GPS and accelerometer indicates that there is possible to make dynamical model of gondola-balloon system and atmospheric dynamics. The analysis deserves another publication which is currently elaborated.

The analysis of the biological part of the experiment is currently in progress, however no living bacteria and fungi were detected.

The camera was tested under stratospheric condition. It worked after the landing and the film from the mission is available at \cite{movie}.

%%%%%%%%%%%%%%%%%%%%%%%%%%%%%%%%%%%%%%%%%%%%%%%%%%%%%%%%%%%%%%%%%%%%%%%%%%%%%%%%%%%%%%%%%%%%%%%%%%%%%%%%%%%%%%%%%%%%%%%%%
\section{Conclusions}
%%%%%%%%%%%%%%%%%%%%%%%%%%%%%%%%%%%%%%%%%%%%%%%%%%%%%%%%%%%%%%%%%%%%%%%%%%%%%%%%%%%%%%%%%%%%%%%%%%%%%%%%%%%%%%%%%%%%%%%%%
The experiment described in the paper was intended as the proof of concept for the stratospheric ELF missions. Although no Schumann resonance was registered the analysis of the results allows to redesign system for future missions. The concepts used in this experiment can be adapted to future similar balloon missions on Mars.

%%%%%%%%%%%%%%%%%%%%%%%%%%%%%%%%%%%%%%%%%%%%%%%%%%%%%%%%%%%%%%%%%%%%%%%%%%%
%Acknowledgments
%%%%%%%%%%%%%%%%%%%%%%%%%%%%%%%%%%%%%%%%%%%%%%%%%%%%%%%%%%%%%%%%%%%%%%%%%%%
\section*{Acknowledgements}
The experiment was funded by the Faculty of Physics, Mathematics and Computer Science of Tadeusz Ko\'{s}ciuszko Cracow University of Technology. We would like to thanks Modular Analog Research Station M.A.R.S., Astronomia Nova organizations and YI.pl company for support. Last but not least, we would like to thank the Hackerspace group in retrieving the gondola after landing.

%%%%%%%%%%%%%%%%%%%%%%%%%%%%%%%%%%%%%%%%%%%%%%%%%%%%%%%%%%%%%%%%%%%%%%%%%%%

%%%%%%%%%%%%%%%%%%%%%%%%%%%%%%%%%%%%%%%%%%%%%%%%%%%%%%%%%%%%%%%%%%%%%%%%%
%koniec bibliografii
%%%%%%%%%%%%%%%%%%%%%%%%%%%%%%%%%%%%%%%%%%%%%%%%%%%%%%%%%%%%%%%%%%%%%%%%%%
%%%%%%%%%%%%%%%%%%%%%%%%%%%%%%%%%%%%%%%%%%%%%%%%%%%%%%%%%%%%%%%%%%%%%%%%%%%

%KONIEC
%%%%%%%%%%%%%%%%%%%%%%%%%%%%%%%%%%%%%%%%%%%%%%%%%%%%%%%%%%%%%%%%%%%%%%%%%


\begin{thebibliography}{99}
 
 \bibitem{ElectricNatureOfStorm}
  D. R. MacGorman, W. D. Rust \emph{The electrical nature of storms} Oxford University Press (1998)
 
   \bibitem{SchumannMeasurment}
     W. O. Schumann, H. König, \emph{\"{U}ber die Beobactung von Atmospherics bei geringsten Frequenzen}. Naturwissenschaften. 41 (8) 183--184 (1954); DOI:10.1007/BF00638174
 
  \bibitem{MARS}
  G. J. Molina-Cuberos, J. A. Morente, B. P. Besser, J. Port\'{\i}, H. Lichtenegger, K. Schwingenschuh, A. Salinas, J. Margineda, \emph{Schumann resonances as a tool to study the lower ionospheric structure of Mars},  Radio Science  41 01 (2006)l DOI:  10.1029/2004RS003187
  
  \bibitem{MARS2}
   J. Kozakiewicz, A. Ku\l{}ak, J. Kubisz, K. Zietara, \emph{Extremely Low Frequency Electromagnetic Investigation on Mars}, Earth, Moon, and Planets, 118, 2, 103--115 (2016)

  \bibitem{BiologicalSystemsELF}
  J. L. Kirschvink, \emph{Comment on "Constraints on biological effects of weak extremely-low-frequency electromagnetic fields"}, Phys. Rev. A 46 4 (1992); 
  
  \bibitem{BaloonSchumann}
  R. Benda et al.,  \emph{Measurements of atmospheric electrical parameters and ELF electromagnetic emissions during a meteorological balloon flight}, Geophysical Research Abstracts Vol. 18, EGU2016-7725 (2016), EGU General Assembly 2016
  
  \bibitem{AntennaTheory}
  C. A. Balanis, \emph{Modern Antenna Handbook}, Wiley-Interscience; 1st edition (2008)

  \bibitem{ELFAntenna}
  A. Ku\l{}ak, J. Knapik, J. Kubisz \emph{Przeno\'{s}na stacja do obserwacji widm dynamicznych p\'{o}l elektromagnetycznych w zakresie ULF} [in Polish], in Proceedings of the VII National Symposium of Radio Science (URSI) (1996)
  
  \bibitem{ELFAntenna2}
  A. Ku\l{}ak, J. Kubisz, S. Klucjasz, A. Michalec, J. M\l{}ynarczyk, Z. Nieckarz, M. Ostrowski, S. Zieba, \emph{Extremely low frequency electromagnetic field measurements at the Hylaty station and methodology of signal analysis},  RADIO SCIENCE 49 6 361--370 (2014); DOI: 10.1002/2014RS005400  
  
  \bibitem{ELFAKulak1}
  A. Kulak, J. Kubisz, A. Michalec et al., \emph{Solar variations in extremely low frequency propagation parameters: I. A two-dimensional telegraph equation (TDTE) model of ELF propagation and fundamental parameters of Schumann resonances}, J. Geophys. Res., 108(A7), 1270, (2003); DOI:10.1029/2002JA009304
  
  \bibitem{ELFAKulak2}
  A. Kulak, J. Mlynarczyk, S. Zieba et al., \emph{Studies of ELF propagation in the spherical shell cavity using a field decomposition method based on asymmetry of Schumann resonance curves},  JOURNAL OF GEOPHYSICAL RESEARCH-SPACE PHYSICS, 111  A10  A10304 (2006); DOI: 10.1029/2005JA011429 
  
  \bibitem{ELFAKulak3}
  Z. Nieckarz,A. Kulak, S. Zieba et al., \emph{Comparison of global storm activity rate calculated from Schumann resonance background components to electric field intensity E-OZ}, 
   ATMOSPHERIC RESEARCH 91 2-4 184--187 SI (2009); DOI: 10.1016/j.atmosres.2008.06.006
 
  \bibitem{ELFAKulak4}
  Z. Nieckarz, S. Zieba, A. Kulak et al., \emph{Study of the Periodicities of Lightning Activity in Three Main Thunderstorm Centers Based on Schumann Resonance Measurements},  MONTHLY WEATHER REVIEW, 137 12 4401--4409 (2009); DOI: 10.1175/2009MWR2920.1
  
  \bibitem{predykcja}
  \url{http://predict.habhub.org/}
  
  \bibitem{GOOGLEEarth}
  \url{https://www.google.com/earth/}
    
  \bibitem{gpsVisualizer}  
  \url{http://www.gpsvisualizer.com/}  
    
   \bibitem{movie} 
   Movie from the mission:
   \url{https://www.youtube.com/watch?v=3Cyj75oyoII}
   
   \bibitem{AdditionalMaterials}
   Additional material from the mission are available at \url{http://fizyk.ifpk.pk.edu.pl/~rkycia/groups/elfHunters.html}
    
\end{thebibliography}
\end{document}